\documentclass  [12pt] {article}
\def\bbbone{{\mathchoice {\rm 1\mskip-4mu l} {\rm 1\mskip-4mu l}
{\rm 1\mskip-4.5mu l} {\rm 1\mskip-5mu l}}}
\def\vp{\vec{p}\, {}^2}

\newcommand{\dd}[2]{\frac{\partial #1}{\partial #2}}
\begin{document}

\title{%
Doubly Special Relativity  and de Sitter space }
\author{ Jerzy {Kowalski--Glikman}\thanks{email jurekk@ift.uni.wroc.pl}~\thanks{Research  partially supported
by the    KBN grant 5PO3B05620.}~ and Sebastian Nowak\thanks{email
pantera@ift.uni.wroc.pl}\\  \\ {\em Institute for Theoretical
Physics}\\ {\em University of Wroc\l{}aw}\\ {\em Pl.\ Maxa Borna 9}\\
{\em Pl--50-204 Wroc\l{}aw, Poland}} \maketitle

\begin{abstract}
In this paper we recall the construction of Doubly Special Relativity (DSR)  as
a theory with energy-momentum space being the four dimensional de Sitter space.
Then the bases  of the DSR theory  can be understood as different coordinate
systems on this space. We investigate the emerging geometrical picture of
Doubly Special Relativity by presenting the basis independent features of DSR
that include the non-commutative structure of space-time and the  phase space
algebra. Next we investigate the relation between our geometric formulation and
the one based on quantum  $\kappa$-deformations of the Poincar\'e algebra.
Finally we re-derive the five-dimensional differential calculus using the
geometric method, and use it to write down the deformed Klein-Gordon equation
and to analyze its plane wave solutions.
\end{abstract}

\section{Introduction}

Without doubts the quest for a theory of quantum gravity is one of the most
important  challenges of the high energy physics. One of the stumbling blocks
on our way to formulate and understand quantum gravity was notorious lack of
any kind of experimental tests and evidences. In the recent years however the
developments on this field have undertaken a sudden turn. Contrary to the
earlier expectations it turns out that we are likely to being close to discover
``quantum gravity signals'' in experiments, if we do not see them already in
the anomalous behavior of ultra-high-energy cosmic rays and TeV photons (for up
to date review of the ``quantum gravity phenomenology'' see
e.g.~\cite{Amelino-Camelia:2002vw} and references therein.)

These developments have led to the opening of  new and rapidly growing field of
research. Since the observable effects are due not to the extremely strong
gravitational fields but rather to cumulation of minute effects one should look
for the traces of quantum gravity in weak gravitational field regime, something
that might be called ``quantum special relativity'', being a limit of quantum
gravity in a similar way Special Relativity is a limit of General Relativity.
Doubly Special Relativity\footnote{Let us note at this point that some authors
\cite{Ahluwalia:2002ye} criticized the use of the term ``doubly special
relativity'', arguing that it would be more correct to call the theory we are
considering ``special relativity with two invariant scales''. Though we fully
agree with the arguments presented there, and even given the fact that the term
``deformed special relativity'' (with the same acronym, DSR) would perhaps
better describe the nature of the DSR proposal, for historical reasons we
decide to follow the original term proposed by  Amelino-Camelia.}
\cite{Amelino-Camelia:2000ge}, \cite{Amelino-Camelia:2000mn} is a proposal of
such a theory. The idea is that there exist in nature two observer-independent
scales, of velocity, identified with the speed of light, and of mass, which is
expected to be of order of Planck mass\footnote{It was observed by many authors
that the ``trans-Planckian'' physics may require introduction of an additional
scale, identified with the Planck mass or Planck length (see \cite{add} for
more details and early references.) Let us note also that in the recent
literature there appear proposals to introduce another second scale, which is
interpreted physically as maximal acceleration \cite{Schuller:2002fn},
\cite{KalyanaRama:2002qh}, \cite{Feoli:2002dv}. It is not clear to us if and
how such  theoretical models are related to DSR. }.

The appearance of the second scale is to be understood as a trace of quantum
structure of space-time, which is present even if the local degrees of freedom
of gravitational field are switched off. Some heuristic arguments supporting
such an interpretation have appeared in \cite{Ahluwalia:2002ye}. Recently
 it was shown that  2+1 gravity
provides an example of a DSR theory \cite{Amelino-Camelia:2003xp},
\cite{Freidel:2003sp}. This opens the possibility to interpret  DSR  as a
semi-classical, flat space limit of quantum gravity in $3+1$ dimensions, as
follows. It is well known that in the limit $G \rightarrow 0$ general
relativity becomes a topological field theory, whose only solution (without
sources) is flat space-time. It is therefore possible that when one takes an
appropriate semiclassical, weak coupling limit of quantum gravity $G
\rightarrow 0$, $\hbar \rightarrow 0$, one finds that $\kappa^{-1} \equiv
\lim_{G,\hbar \rightarrow 0}\, \sqrt{G/\hbar}$ remains finite (and equal to the
Planck length.) By construction $\kappa$ would be a new fundamental,
observer-independent scale governing the particle kinematics in flat space time
at ultra high energy. The analysis of particle kinematics in 2+1 gravity
supports also the claim that symmetries of particle kinematics are to be
described by a quantum group, and that   particles are labelled by
representations of this quantum group.

Introduction of the second observer-independent scale makes it necessary, of
course, to modify the rules that govern transformations from one inertial
observer to another. In the algebraic language this is equivalent to replacing
the standard Poincar\'e algebra of Special Relativity by another, deformed
algebra, tending to the former in the limit of large mass scale
$\kappa\rightarrow\infty$. Indeed such an example of Doubly Special Relativity
has been found and described in some details in \cite{jkgminl},
\cite{rbgacjkg}. It has been shown further \cite{juse}, \cite{lunoDSR} that the
structure of this theory, called nowadays DSR1 or bicrossproduct (basis) DSR,
with non-commutative space-time, is identical with the quantum deformations of
Poincar\'e symmetry, known as $\kappa$-Poincar\'e algebra \cite{kappaP},
\cite{kappaM}.

 DSR1 is of course not the only Doubly Special Relativity. Another theory of
this kind has been constructed by Magueijo and Smolin \cite{Magueijo:2001cr},
and it was this paper that became a main motivation to investigate the whole
class of DSR theories, instead of particular examples. Such investigations have
been undertaken in  \cite{juse}, \cite{Kowalski-Glikman:2002jr} and the results
of these papers can be summarized as follows:
\begin{enumerate}
\item There exists a whole class of Doubly Special Relativity Theories, for whose the Lorentz
symmetry algebra is not deformed. This means that, contrary to the statements
one can sometimes find in the literature, in DSR theories we do not have to do
with Lorentz symmetry breaking, i.e., there exist a subgroup of the group of
symmetries of the theory, whose algebra is exactly the Lorentz algebra of
Special Relativity.
\item The algebra of Lorentz transformations and momenta is deformed, though.
 All such deformations of Poincar\'e algebra (i.e., all DSR
theories) are related to each other by reparametrizations of momentum variables
\cite{juse}, \cite{Kowalski-Glikman:2002jr}. In this paper, for illustrative
purposes we will employ the DSR1 theory, described by the $\kappa$-Poincar\'e
algebra in the bicrossproduct basis \cite{kappaP}, \cite{kappaM}.
\item Each of the deformed algebras of Lorentz transformation and momenta can be provided with the
quantum algebra  structure, in particular with the one identical with
$\kappa$-Poincar\'e. This structure makes it possible in turn to construct the
phase space algebra \cite{crossalg}, \cite{luno}, i.e., the set of commutators
between Lorentz generators, momenta and positions. It turns out also that
positions do not commute, with either $\kappa$-Minkowski \cite{kappaM}, or
Snyder \cite{snyder} type of non-commutativity, so that in DSR we have to do
with the Lie type, non-commutative space-time structure \cite{lunoDSR},
\cite{Kowalski-Glikman:2002jr}.
\item Since DSR is to be a theory of particle kinematics, one cannot restrict oneself to
the energy momentum space only (as it was implicit at the early stages of the
development of DSR programme.) Instead a DSR theory is defined by the phase
space algebra, which in addition to the energy momentum sector contains the
noncommutative space-time structure, and a  set of nontrivial cross commutators
of energy/momenta with space/time positions.
\end{enumerate}

It has been shown in the recent paper \cite{Kowalski-Glikman:2002ft} that all
these properties can be understood if one employs a simple geometrical
language. Namely, any DSR theory can be understood as a particular coordinate
system on four dimensional de Sitter space of momenta imbedded in five
dimensional Minkowski space. In this picture the Lorentz transformations are
identified with the $SO(3,1)$ subgroup of the $SO(4,1)$ group of the symmetries
of de Sitter space, while the (non-commutative) positions with the remaining
four generators, being ``translations'' in energy-momentum space. The present
paper is devoted to investigations of the geometric picture associated with DSR
theories in more details. In Section 2 we present some basic information
concerning de Sitter space of momenta and its symmetries, in Section 3 we
discuss the geometric picture of the DSR theories in more details. Section 4 is
devoted to investigating relation between geometric de Sitter space picture and
the quantum algebraic approach to DSR. In Section 5 we describe the
construction of the covariant differential calculus, and use it to write down
the deformed Klein-Gordon equation as well as to derive the plane wave
solutions of the this equation. Section 6 is devoted to explaining relation
between  de Sitter space formalism and some other techniques employed in
investigations of DSR theories.

Throughout this paper we  try to formulate some results of general nature. For
the illustrative purposes however we employ the  DSR1 theory. In this theory
the (commutative) momenta $p_\mu$ transform under action of boosts $N_i$, and
rotations $M_i$  as follows \cite{kappaM}
\begin{equation}\label{c}
[N_i, p_j] =  i\,  \delta_{ij}\,
 \left(  {\kappa\over 2} \left( 1-e^{- 2{p_{0}/ \kappa}}
 \right) + {1\over 2\kappa} \vec{p}\,{}^{ 2}\, \right) - i\,
{1\over \kappa} p_{i}p_{j} ,\quad [N_i, p_0] = i p_i,
\end{equation}
\begin{equation}\label{d}
 [M_i, p_j] = i\, \epsilon_{ijk}\, p_k, \quad [M_i, p_0] = 0.
\end{equation}
There are of course many other DSR theories, and we list the basic properties
of some of them in the Appendix.

\section{De Sitter space of momenta}

Doubly Special Relativity was initially formulated
\cite{Amelino-Camelia:2000ge}, \cite{Amelino-Camelia:2000mn}, \cite{jkgminl},
\cite{rbgacjkg} as a theory of nonlinear realization of the Lorentz symmetry on
the energy-momentum space. It has been obvious however that one should extend
it to the formulation based on the complete phase space. This has been done in
\cite{Kowalski-Glikman:2002jr} with the help of the quantum $\kappa$-Poincar\'e
algebra structure and the so-called Heisenberg double procedure
\cite{crossalg}, \cite{luno}. In the paper \cite{Kowalski-Glikman:2002ft} it
has been shown that an equivalent, geometric method of deriving the phase space
algebra exists. This method is based on appropriate interpretation of ten
symmetries of the four dimensional space of momenta, which contrary to the
standard case is assumed to be a maximally symmetric space of constant positive
curvature, the de Sitter space. Let us now turn to the description of this
space and its symmetries.

Consider the five dimensional space of variables $\eta_A$, $A=0,\ldots,4$ of
dimension of momentum equipped with the Minkowski metric
\begin{equation}\label{1}
ds^2 = g^{AB}d\eta_A d\eta_B = -d\eta_0^2 + d\eta_i^2 + d\eta_4^2.
\end{equation}
This metric is invariant under the action of the group $SO(4,1)$, whose
generators are rotations $M_i$, boosts $N_i$, and four remaining generators,
which we will denote $X_\mu$ and call ``positions'' (this terminology will
become clear later on.) The Lorentz generators form the ${\cal SO}(3,1)$
subalgebra of the ${\cal SO}(4,1)$   algebra and satisfy
$$
[M_i, M_j] = i\, \epsilon_{ijk} M_k, \quad [M_i, N_j] = i\, \epsilon_{ijk} N_k,
$$
\begin{equation}\label{2}
  [N_i, N_j] = -i\, \epsilon_{ijk} M_k,
\end{equation}
These operators act on the first four variables $\eta_\mu$ in a standard way,
to wit
\begin{equation}\label{3}
  [M_i, \eta_j] = i\epsilon_{ijk}\eta_k, \quad [N_i, \eta_j] = i\, \delta_{ij}\, \eta_0, \quad [N_i, \eta_0] = i\,  \eta_i,
\end{equation}
and leave the variable $\eta_4$ invariant. The commutational relations between
these generators and positions are defined by  decomposing  ${\cal SO}(4,1)$
into ${\cal SO}(3,1)$ and the remainder. In the case of Cartan decomposition,
we have the relations
\begin{equation}\label{4}
 [X_0, X_i] = -\frac{i}{\kappa^2}\, N_i, \quad [X_i, X_j] = \frac{i}{\kappa^2}\,\epsilon_{ijk}\, M_k
\end{equation}
We will identify $X_\mu$ with position operators (i.e.\ generators of
``translations'' of momenta,) and for that reason we have re-scaled them so
that they have the dimension of length. Then one easily sees that the relations
(\ref{4}) correspond to space-time non-commutativity of Snyder's type
\cite{snyder}.

Another natural decomposition is possible, leading to the commutators
\begin{equation}\label{5}
  [x_0, x_i] = -\frac{i}{\kappa}\, x_i, \quad [x_i, x_j] = 0.,
\end{equation}
This quite different type of non-commutativity corresponds to the
$\kappa$-Minkowski space-time \cite{kappaM}. Note  that contrary to (\ref{4}),
eqs.\ (\ref{5}) describe the Lie-type non-commutativity of space-time. To
emphasize this difference we will denote positions in Snyder's non-commutative
space-time by capital letters, and those in $\kappa$-Minkowski space-time by
small ones. The remaining ${\cal SO}(4,1)$ commutators read then
\begin{equation}\label{6a}
 [N_i, X_0] = i X_i, \quad [N_i, X_j] = i\, \delta_{ij}\, X_0
\end{equation}
and
\begin{equation}\label{6b}
 [N_i, x_0] = i x_i - \frac{i}\kappa\,  N_i, \quad [N_i, x_j]=i \delta_{ij} x_0 - \frac{i}\kappa\, \epsilon_{ijk} M_k.
\end{equation}
The relation between $\kappa$-Minkowski and Snyder's position variables $x_\mu$
and $X_\mu$ reads
\begin{equation}\label{7}
 x_0 = X_0, \quad x_i = X_i + \frac1\kappa N_i,
\end{equation}
so that indeed these variables are related by simple rearrangement of the basis
of the Lie algebra ${\cal SO}(4,1)$ of symmetries of de Sitter space.

 Knowing the algebra, we can write down the action of ${\cal SO}(4,1)$
generators on the variables $\eta_A$. The Lorentz boosts and rotations act in
the standard way
\begin{equation}\label{8}
 [M_i, \eta_j] = i\epsilon_{ijk}\, \eta_k, \quad [M_i, \eta_0]=[M_i, \eta_4] =0
\end{equation}
\begin{equation}\label{9}
 [N_i,\eta_j] = i\, \delta_{ij}\, \eta_0, \quad [N_i, \eta_0] = i \eta_i,\quad [N_i, \eta_4] =0
\end{equation}
The cross commutators of $\kappa$-Minkowski positions and $\eta_A$ have the
form
\begin{equation}\label{10}
  [x_0,\eta_4] = \frac{i}\kappa\, \eta_0, \quad [x_0,\eta_0] = \frac{i}\kappa\, \eta_4, \quad [x_0,\eta_i] = 0,
\end{equation}
\begin{equation}\label{11}
  [x_i, \eta_4] = [x_i, \eta_0] =\frac{i}\kappa\, \eta_i, \quad [x_i, \eta_j] = \frac{i}\kappa\,
\delta_{ij}(\eta_0 - \eta_4),
\end{equation}
while for the Snyder's position variables one finds
\begin{equation}\label{12}
  [X_0,\eta_4] = \frac{i}\kappa\, \eta_0, \quad [X_0,\eta_0] = \frac{i}\kappa\, \eta_4, \quad [X_0,\eta_i] = 0,
\end{equation}
\begin{equation}\label{13}
  [X_i, \eta_4] = \frac{i}\kappa\, \eta_i,\quad  [X_i, \eta_0] =0
  \quad [X_i, \eta_j] = -\frac{i}\kappa\,
\delta_{ij}  \eta_4,
\end{equation}
As it will be shown in the next section equations (\ref{8})--(\ref{13}) can be
used  to reconstruct the whole of the phase space of any particular DSR theory.

\section{DSR theories as coordinates on de Sitter space}
It is well known that the algebra ${\cal SO}(4,1)$ discussed in the previous
section is an algebra of symmetries of de Sitter space defined as a following
surface in the five-dimensional Euclidean space of Minkowski signature
(\ref{1})
\begin{equation}\label{14}
 -\eta_0^2 + \eta_1^2+ \eta_2^2+ \eta_3^2+ \eta_4^2 =\kappa^2,
\end{equation}
Let us define an origin ${\cal O}$ of de Sitter space as a point invariant
under action of the $SO(3,1)$ subgroup of this symmetry group, cf.~(\ref{8}),
(\ref{9}); thus the point ${\cal O}$ has the coordinates $(0,0,0,0,\kappa)$.
Let us define coordinates $p_0, p_i$ (physical momenta) on the surface
(\ref{14}) by
\begin{equation}\label{15}
 \eta_0 = \eta_0(p_0, \vec{p}\, {}^2), \quad \eta_i =  p_i\, \eta(p_0, \vec{p}\, {}^2),
 \quad \eta_4 = \sqrt{\kappa^2 +\eta_0^2 - \eta_1^2- \eta_2^2- \eta_3^2},
\end{equation}
where we explicitly assumed that the coordinates $p_0, p_i$ transform under
rotations in the standard way, and such that $\eta_0(0,\vec{0}) =0$. Thus the
origin ${\cal O}$ corresponds to the zero value of the physical momenta.

Of course, the coordinates $p_0, p_i$ are defined only up to an arbitrary
redefinition, i.e., general coordinate transformation, leaving the origin
${\cal O}$ invariant, and may or may not cover the whole of the space
(\ref{14}).

As an example take the following coordinates
\begin{eqnarray}
{\eta_0} &=& \kappa\, \sinh \frac{p_0}\kappa + \frac{\vec{p}\,{}^2}{2\kappa}\,
e^{  \frac{p_0}\kappa} \nonumber\\
\eta_i &=&   p_i \, e^{  \frac{p_0}\kappa} \nonumber\\
{\eta_4} &=&  \kappa\, \cosh \frac{p_0}\kappa  - \frac{\vec{p}\,{}^2}{2\kappa}
\, e^{  \frac{p_0}\kappa}.   \label{16}
\end{eqnarray}
Using the expressions (\ref{9}) and  Leibniz rule one easily finds that these
coordinates correspond to the transformation rules of DSR in the bicrossproduct
basis (\ref{c}). Thus the prescription (\ref{16}) provides the geometric
definition of the  DSR1. Similar prescriptions are possible for other DSR
models, and we present them in the Appendix.
\newline

A natural question arises as to whether there is a freedom in choosing the
coordinates on de Sitter space, corresponding to the given DSR transformation
rules, i.e., given commutators of Lorentz boosts and physical momenta of a
particular DSR model. To answer it, let us consider the most general DSR boost
transformations (assuming, as usual, the standard action of rotations)
\begin{equation}\label{17}
[N_i, p_j] = i\, \delta_{ij} \alpha + i\, p_i p_j \beta
\end{equation}
(the term of form $\sim \epsilon_{ijk} p_k $ is excluded by Jacobi identity)
and
\begin{equation}\label{18}
[N_i, p_0] = i\, p_i\, \gamma,
\end{equation}
where $\alpha, \beta, \gamma$ are functions of $p_0$, $\vp$ only, i.e., they
are  scalars with respect to rotations and  are restricted by Jacobi identity
\begin{equation}\label{jac}
 \dd{\alpha}{p_0}\, \gamma + 2 \dd{\alpha}{\vp}\, (\alpha + \vp\beta) - \alpha\beta  = 1
\end{equation}
 Take now some $\eta_\mu$ transforming according to
(\ref{9}).  Because of the undeformed action of rotations with no loss of
generality we can make the ansatz
\begin{equation}\label{19a}
 \eta_0 = \eta_0(p_0, \vp), \quad \eta_i = p_i\, \eta(p_0, \vp)
\end{equation}
Assume now that the momenta of (\ref{19a}) transform under boost according to
(\ref{17}), (\ref{18}). Using (\ref{9}) and  Leibniz rule we get the following
system of differential equations
\begin{equation}\label{19}
\dd{\eta_0}{p_0}\, [N_i, p_0] + 2 \dd{\eta_0}{\vp} p_j\, [N_i, p_j] = i p_i\,
\eta
\end{equation}
\begin{equation}\label{20}
[N_i, p_j]\, \eta + \dd{\eta}{p_0}\, [N_i, p_0]\, p_j+ 2 \dd{\eta}{\vp} p_j\,
p_k\, [N_i, p_k] = i\, \delta_{ij} \eta_0.
\end{equation}
From eq.~(\ref{20}) we have
\begin{equation}\label{21}
  \alpha\, \eta =\eta_0
\end{equation}
and
\begin{equation}\label{22}
  \beta\, \eta +\dd{\eta}{p_0}\,\gamma +2 \dd{\eta}{\vp}\left(\alpha + \vp\, \beta\right) =0
\end{equation}
It can be checked that for any solution of eqs.~(\ref{21}), (\ref{22})
eq.~(\ref{19}) is satisfied identically due to the Jacobi identity (\ref{jac}).
\newline

As an example consider again the DSR1, for whose
$$
\alpha = \frac{\kappa}{2}\left(1 - e^{-2p_0/\kappa}\right) +
\frac{\vp}{2\kappa}, \quad \beta = -\frac1\kappa, \quad \gamma =1.
$$
Eq.~(\ref{22}) can be then solved explicitly, giving
\begin{equation}\label{23}
 \eta = e^{p_0/\kappa} \, f\left[\kappa\cosh\left(\frac{p_0}\kappa\right) - \frac{\vp}{2\kappa}\, e^{p_0/\kappa}\right],
\end{equation}
where $f$ is an arbitrary function of the Casimir ${\cal C}$ (\ref{24a}). In
particular, for $f=1$ we get the expression (\ref{16}).

Let us observe that the argument of the function $f$ above is
nothing but $\eta_4$ in (\ref{16}) and this suggests that the
general solution of eq.~(\ref{22}) is of the form
\begin{equation}\label{24}
  \eta(p_0, \vp) = A(p_0, \vp) \, f({\cal C}),
\end{equation}
where $A(p_0, \vp)$ is a particular solution of eq.\ (\ref{22})
($A(p_0, \vp)=e^{p_0/\kappa}$ for DSR1), and
\begin{equation}\label{24a}
  {\cal C} \equiv \eta_0^2 - \vec{\eta}\, {}^2 = \left(2\kappa \sinh \left(\frac{p_0}{2\kappa}\right)\right)^2 -
\vec{p}\,{}^2\, e^{p_0/\kappa}= m^2
\end{equation}
   is the Casimir of the algebra (\ref{17}), (\ref{18}). It can
be easily shown that any solution of eq.~(\ref{22}) is of this
form. However we have not be able to prove that any solution of
this equation has the form (\ref{24}) in general case, i.e., for
arbitrary choice of $\alpha$, $\beta$, and $\gamma$.
\newline

Let us now turn to eqs.~(\ref{10}), (\ref{11}) in order to get the remaining
commutators of the phase space. Let us first note that
\begin{equation}\label{25}
 \eta_4 = \sqrt{\kappa^2 +\eta_0^2 - \vec{\eta}\,{}^2} = \sqrt{\kappa^2 +\eta^2(\alpha^2 - \vp)}
\end{equation}
Then one plugs this formula along with the expressions $\eta_0 = \alpha\,
\eta$, $\eta_i = p_i\, \eta$ to eqs.~(\ref{10}), (\ref{11}) and reads the phase
space commutators using  Leibniz rule. For example, for the bicrossproduct
basis with $f=1$, one gets
$$ [p_0,
x_0] = -i, \quad [p_i, x_0] =  \frac{i}\kappa\, p_i,
$$
\begin{equation}\label{26}
[p_i, x_j] = i\, \delta_{ij} \, e^{-
2p_0/\kappa} -\frac{i}{\kappa^2}\left(\vec{p}\,{}^{ 2}\, \delta_{ij} - 2
p_{i}p_{j}\right), \quad [p_0, x_i] = -\frac{2i}\kappa\, p_i
\end{equation}

Let us note that in addition to the freedom of choosing de Sitter coordinates
corresponding to given Lorentz transformation rules, there exists an additional
freedom in construction of the phase space. Indeed it is clear from
eq.~(\ref{9}) that the action of Lorentz generators is unaffected by the
replacement $\eta_\mu \rightarrow -\eta_\mu$, $\eta_4 \rightarrow \eta_4$,
while such replacement clearly changes the position--momentum commutators. For
example if we replace the bicrossproduct basis above with
\begin{eqnarray}
{\eta_0} &=& -\kappa\, \sinh \frac{P_0}\kappa - \frac{\vec{P}\,{}^2}{2\kappa}\,
e^{  \frac{P_0}\kappa} \nonumber\\
\eta_i &=&   -P_i \, e^{  \frac{P_0}\kappa} \nonumber\\
{\eta_4} &=&  \kappa\, \cosh \frac{P_0}\kappa  - \frac{\vec{P}\,{}^2}{2\kappa}
\, e^{  \frac{P_0}\kappa},   \label{26a}
\end{eqnarray}
(which corresponds to $f=-1$ in (\ref{24}),) and make use of
eqs.~(\ref{10}), (\ref{11}), we find
\begin{equation}\label{26b}
  [P_0, x_0] = i, \quad [P_i, x_0] =  -\frac{i}\kappa\, P_i, \quad [P_i, x_j] =-
i\, \delta_{ij} , \quad [P_0, x_i] =0.
\end{equation}
This is nothing but the $\kappa$-Minkowski phase space, well known from the
literature on quantum $\kappa$-Poincar\'e (quantum) algebra and
$\kappa$-Poincar\'e (quantum) group (see, e.g., \cite{kappaM}.) Let us
therefore investigate the relation between quantum algebraic and geometric
approaches in more details.

\section{De Sitter vs.~quantum algebraic approach to DSR}

The phase space of DSR has been first derived not by m aking use of  the
geometric picture presented in the preceding section, but in the framework of
quantum algebras. Let us recall this construction.

The starting point is the extension of the algebra (\ref{c}), (\ref{d}) to the
Hopf algebra, by introducing additional structures: co-product $\Delta$ and
antipode $S$, as follows \cite{kappaP}
\begin{eqnarray} %{l}
\displaystyle
 && \Delta(M_{i}) = M_{i}\otimes \bbbone + \bbbone
\otimes M_{i}\, , \cr\cr \displaystyle && \Delta(N_{i}) = N_{i}\otimes
e^{-{P_{0}/ \kappa}} + \bbbone \otimes N_{i} - {1\over \kappa}
\epsilon_{ijk}M_{j}\otimes P_{k}\, , \cr\cr \displaystyle && \Delta(P_{i}) =
P_{i}\otimes \bbbone + e^{-{P_{0}/ \kappa}} \otimes P_{i}\, , \cr\cr
\displaystyle && \Delta(P_{0}) = P_{0}\otimes \bbbone +  \bbbone \otimes
P_{0}\, , \label{27}
\end{eqnarray}

%%%%

\begin{eqnarray} & S(P_{i}) =e^{-{P_{0}\over \kappa}} P_{i} \quad & S(P_{0}) = -
P_{0} \cr\cr & S(M_{i}) = - M_{i} \quad &S(N_{i}) = - e^{{P_{0}\over \kappa}}
N_{i} + {1\over \kappa} \varepsilon_{ijk} e^{{P_{0 }\over \kappa}} P_{j}M_{k}\,
.\label{28}
\end{eqnarray}

In equations above we used capital letter $P_\mu$ to denote momenta. The
difference between these variables and the variables $p_\mu$ used in the
previous section will be explained in a moment.

As explained in \cite{crossalg}, \cite{luno}, \cite{Kowalski-Glikman:2002jr}
the co-product encodes information concerning the phase space of the theory. To
disclose this information one uses the procedure called ``Heisenberg double'',
which consists of the following:

\begin{enumerate}
\item One defines the bracket $<\star, \star>$ between momentum variables $P,Q$ and position variables $X,Y$ in
a natural way as follows
\begin{equation}\label{16a}
 <P_\mu, x_\nu> =  -i \eta_{\mu\nu}, \quad \eta_{\mu\nu} = \mbox{diag}(-1,1,1,1).
\end{equation}
\item This bracket is to be consistent with the co-product structure in the following sense
$$
 <P, xy> = <P_{(1)}, x><P_{(2)}, y>,
 $$
 \begin{equation}\label{16b}
 <PQ,x> =<P, x_{(1)}><Q_{(2)}, x_{(2)}>,
\end{equation}
where we use the natural (``Sweedler'') notation for co-product $$\Delta {\cal
T} = \sum {\cal T}_{(1)} \otimes {\cal T}_{(2)}.$$ It should be also  noted
that by definition
$$<\bbbone, \bbbone> =1.$$ One sees immediately that the fact that momenta
commute translates to the fact that positions co-commute
\begin{equation}\label{16c}
  \Delta x_\mu = \bbbone \otimes x_\mu + x_\mu \otimes \bbbone.
\end{equation}
Then the first equation in (\ref{16b}) along with (\ref{16a}) can be used to
deduce the form of the space-time commutators.
\item  It remains only to derive the cross relations between momenta and positions. These can be found
from the definition of the so-called Heisenberg double (see \cite{crossalg})
and read
\begin{equation}\label{16d}
 [P,x] =  x_{(1)}<P_{(1)}, x_{(2)}>P_{(2)}-xP
\end{equation}
\end{enumerate}

As an example let us perform these steps in the DSR1
%\cite{maru}, \cite{crossalg}.
It follows from (\ref{16b}) that
$$
<P_i, x_0 x_j> = -\frac1\kappa\, \delta_{ij}, \quad <P_i,  x_jx_0>=0,
$$
from which one gets
\begin{equation}\label{17a}
[x_0, x_i] = -\frac{i}\kappa\, x_i.
\end{equation}
Let us now make use of (\ref{16d}) to get the standard relations
\begin{equation}\label{17b}
[P_0, x_0] = -i, \quad [P_i, x_j] = i \, \delta_{ij}.
\end{equation}
It turns out that this algebra contains one more non-vanishing commutator,
namely
\begin{equation}\label{17c}
 [P_i, x_0] = -\frac{i}\kappa\, P_i.
\end{equation}
Thus starting from the $\kappa$-Poincar\'e co-product (\ref{27})  using the
Heisenberg double prescription one derives the phase space (\ref{26b}).

It is then obvious that the existence of the second phase space
(\ref{26}) must be related to the existence of another co-product
on the same deformed $\kappa$-Poincar\'e algebra (\ref{c}),
(\ref{d}). This co-product can be computed by a rather tedious
procedure and we have been able to construct it only up to linear
terms in the first factor, and therefore we will not present it
here explicitly. Using this co-product and the Heisenberg double
procedure one obtains the phase space relation (\ref{26}).
\newline

Similarly one deduces the commutators of boost an position operators. One
starts with the pairing
\begin{equation}\label{29}
<N_i, x_j> = i \delta_{ij} x_0, \quad <N_i, x_0> = i  x_i.
\end{equation}
Then, after simple computations, making use of  eq.~(\ref{27}) we get
\begin{equation}\label{30}
[N_i, x_j] = i \delta_{ij} x_0 - \frac{i}\kappa\, \epsilon_{ijk} M_k, \quad
[N_i, x_0] = i x_{i} - \frac{i}\kappa\, N_i.
\end{equation}
This relations are of course identical with the ones obtained by geometrical
method (\ref{6b}). One should note at this point that, as proved in
\cite{Kowalski-Glikman:2002jr} the commutators (\ref{30}) are basis
independent, because in the Heisenberg double procedure they picked up
contributions from the first order terms in co-product, that are left unchanged
if one turns from one DSR basis to another. This of course in perfectly
consistent with the universal role the ${\cal SO}(4,1)$ algebra of symmetries
of de Sitter space plays in the geometric formulation of DSR.

\section{Differential calculus and plane waves}

As emphasized above the space-time of DSR theories is a non-commutative space
(with Lie type non-commutativity, contrary to the ``central-charge-type''
non-commutativity considered in string theory.) It is therefore a highly
non-trivial exercise to establish the differential calculus on this space-time.
Such a calculus in turn is necessary to give meaning to derivative and other
differential operators, needed to formulate field theory.

On $\kappa$-Minkowski space-time one can construct two distinct differential
calculi. Chronologically the first one is the calculus bi-covariant under
action of the full DSR algebra; however it is necessarily five dimensional
\cite{5dcalc1}. Such a calculus (in four space-time dimensions) was derived in
\cite{5dcalc2}, \cite{5dcalc3} using the methods of quantum groups theory. Here
we present an alternative (though equivalent) derivation of this form of
differential calculus.

On $\kappa$-Minkowski space-time there also exists
 a four-dimensional translational invariant calculus proposed by Majid and Oeckl
\cite{majid}, which is however not covariant under the action of the full DSR
algebra (generated by momenta and Lorentz generators). For this reason we will
not consider it here. It should be noted that this calculus has been used in
\cite{Amelino-Camelia:2002tc} to derive properties of the plane waves in DSR1
theory.
\newline

The problem we are to solve here is the following. In commutative space-time
positions commute with differentials (one forms). However here we are working
with non-commutative space-time, and thus we cannot assume a priori that
positions commute with one forms. Instead, let us take a basis of one forms.
This basis should include differentials $dx_\mu$, but it turns out that in
order to obtain a consistent covariant differential calculus one must add  one
more one-form, which we will denote $\phi$. Let us therefore denote the
elements of the basis of one forms by $\chi_A = (dx_\mu, \phi)$, $A =0,
\ldots,4$. In the next step we must postulate the form of the commutator
$[x_\mu,\chi_A]$ and we assume that it is proportional to the linear
combination of the basic one-forms, to wit
\begin{equation}\label{31}
[x_\mu,\chi_A] = f^{B}{}_{\mu A}\, \chi_B.
\end{equation}
Since positions form Lie type algebra, $[x_0, x_i] = -\frac{i}\kappa\, x_i$,
 taking the commutator of both sides of this equation with $\chi_A$ and using
the Jacobi identity and (\ref{31}) we get
\begin{equation}\label{32}
f^{B}{}_{0 A}\, f^{C}{}_{i B}-f^{B}{}_{i A}\, f^{C}{}_{0 B} =\frac{i}\kappa\,
f^{C}{}_{i A}.
\end{equation}
Next we apply the exterior derivative to both sides of $[x_0, x_i] =
-\frac{i}\kappa\, x_i$ and we use the Leibnitz rule to obtain
$$
[x_0, dx_i]+[dx_0, x_i]  = -\frac{i}\kappa\, dx_i
$$
that is
\begin{equation}\label{33}
 f^{B}{}_{0i} -f^{B}{}_{i0} =-\frac{i}\kappa\, \delta^B{}_i.
\end{equation}
Similarly, using $[x_i, x_j]=0$ we find
\begin{equation}\label{33a}
f^{B}{}_{ij} -f^{B}{}_{ji} =0.
\end{equation}
Note that by taking exterior derivative once again and using the fact that $d^2
=0$ we find
\begin{equation}\label{33b}
 [dx_\mu, dx_\nu] =0
\end{equation}

 We need to append these conditions with the
covariance requirement, i.e., the condition that both sides of (\ref{31})
transform in the same way under action of rotations and boosts. Using Jacobi
identity we find (cf.\ (\ref{6b})
\begin{equation}\label{d1}
 - [x_0, [\chi_A, M_i]] =f^{B}{}_{0 A}\, [M_i,
\chi_B],
\end{equation}
\begin{equation}\label{d2}
 i\epsilon_{ijk}\, f^{B}{}_{k A}\, \chi_B- [x_j, [\chi_A,
M_i]] =f^{B}{}_{j A}\, [M_i, \chi_B],
\end{equation}
\begin{equation}\label{d3}
 i f^{B}{}_{i A}\, \chi_B+\frac{i}\kappa [\chi_A,N_i]- [x_0,
[\chi_A, N_i]] =f^{B}{}_{0 A}\, [N_i, \chi_B],
\end{equation}
\begin{equation}\label{d4}
 i\delta_{ij}\, f^{B}{}_{0 A}\, \chi_B
+\frac{i}\kappa\,\epsilon_{ijk}\, [\chi_A,M_k]- [x_j, [\chi_A, N_i]]
=f^{B}{}_{j A}\, [N_i, \chi_B],
\end{equation}
 Any solution of eqs.~(\ref{32})  -- (\ref{d4}) determines
the first order covariant calculus of differential forms. It is now clear why
the four dimensional covariant differential calculus does not exist. To see
this recall that the Lorentz generators form together with the positions the
algebra $SO(4,1)$. Eqs.~(\ref{32}) -- (\ref{d4}) together with eqs.~(\ref{3}),
(\ref{5}), (\ref{6b}) would therefore define a Lie algebra of the semidirect
product of $SO(4,1)$ with $R^4$, but such algebra could exist only if $x_\mu$
commutes with $dx_\mu$. Therefore we conclude that the minimal dimension of the
covariant differential calculus equals $5$. The same conclusion can be reached,
of course, by tedious analysis of eqs.~(\ref{32}) --
 (\ref{d4}).

Thus we see that the basis of one-forms of the covariant differential calculus
is indeed $\chi_A = (dx_\mu, \phi)$. Since $\phi$ does not carry the space-time
index,  it must be invariant under action of the Lorentz generators
\begin{equation}\label{f1}
[M_i, \phi] = [N_i,\phi] =0,
\end{equation}
because invariance is the only covariant behavior of scalars. Moreover, since
the action of rotations is
 assumed to be classical, we have
\begin{equation}\label{f2}
[M_i, dx_0] =0, \quad [M_i, dx_j] = i\epsilon_{ijk}\, dx_k.
\end{equation}

Next, since the action of boosts must transform one forms into one forms, we
have that
\begin{equation}\label{f3}
 [N_i, dx_0] = idx_i , \quad [N_i,dx_j] = i \delta_{ij} \, dx_0.
\end{equation}
(In principle one can add a term proportional to $i \delta_{ij} \, \phi$ to the
right hand side of the second equation, but such term can be absorbed into
redefinition of $dx_0$.) Then one can solve eqs.~(\ref{32}) --
 (\ref{d4}) to obtain
$$
[x_\mu, \phi] = \frac i\kappa\, dx_\mu, $$ $$ [x_0, dx_0] = \frac i\kappa\,
\phi, \quad [x_0, dx_i] = 0,
$$
\begin{equation}\label{34}
[x_i, dx_0] = \frac i\kappa\, dx_i, \quad [x_i, dx_j] = \frac i\kappa\,
\delta_{ij} \left( dx_0 - \phi\right).
\end{equation}
Compare now the expressions (\ref{34}) with eqs.~(\ref{10}), (\ref{11}). They
are identical if we substitute $dx_\mu = \eta_\mu$, $\phi = \eta_4$. Thus the
$\eta_A$ variables are nothing but the basis of one forms.
\newline

Knowing the basis of one forms, one can try to understand the meaning of the
differential of  function. To this end, one must define a particular ordering
of polynomials, which we assume to be ``$x_0$-to-the-left'' one, and denote it
by $:\, \ast \, :$. Then such an ordering can be, at least formally, extended
to any analytic function of positions. Given such a function, its differential
is defined to be
\begin{equation}\label{35}
df = \partial_\mu f\, dx^\mu + \partial f\, \phi,
\end{equation}
which in turn defines the left partial derivatives $\partial_\mu, \partial$.
Let us now derive the explicit expression for partial derivatives. Using
equation above, after tedious computations one finds the general expression for
the differential \cite{5dcalc3}
$$
d\, :f(x):\, = \, :\left(\kappa\sin(\left(\frac{\partial_0}\kappa\right) +
\frac{i}{2\kappa}\,  e^{i\partial_0/\kappa}\, \frac{\partial^2}{\partial
x_i\partial x_i}\right)\, f:\,  dx_0+ : e^{i\partial_0/\kappa}\,
\frac{\partial}{\partial x_i}\, f:\,  dx_i
$$
\begin{equation}\label{36}
- i:\left( 1 -\cos\left(\frac{\partial_0}\kappa\right) - \frac12\,
e^{i\partial_0/\kappa}\, \frac{\partial^2}{\partial x_i\partial x_i}\right)\,
f:\, \phi.
\end{equation}
and partial derivatives \cite{5dcalc3}
\begin{equation}\label{37}
\partial_0\, :f:\,= \,:\left(\kappa\sin(\left(\frac{\partial_0}\kappa\right) +
\frac{i}{2\kappa}\,  e^{i\partial_0/\kappa}\, \frac{\partial^2}{\partial
x_i\partial x_i}\right)\, f:
\end{equation}
\begin{equation}\label{38}
  \partial_i\, :f:\,= \,: e^{i\partial_0/\kappa}\,
\frac{\partial}{\partial x_i}\, f:
\end{equation}
\begin{equation}\label{39}
  \partial\, :f:\,= \,-i :\left( 1 -\cos\left(\frac{\partial_0}\kappa\right) - \frac12\,
e^{i\partial_0/\kappa}\, \frac{\partial^2}{\partial x_i\partial x_i}\right)\,
f:
\end{equation}
for ordered functions.

Note the remarkable fact that differential operators in
eqs.~(\ref{36})--(\ref{39}) are just expressions (\ref{16}) or (\ref{26a}) with
momenta replaced by appropriate derivatives. This property can be easily
understood if one notices that for the DSR1 momenta with the phase space
(\ref{26b}) one has \cite{5dcalc3}
\begin{equation}\label{40}
 F(P_\mu) \, :f(x):\, = \, :F\left(i\frac{\partial}{\partial x_\mu}\right)\, f(x):
\end{equation}
We can conclude that  this property distinguishes the DSR1 theory among other
DSR theories, but, of course, one could reach the same conclusion by observing
that the phase space structure (\ref{26b}) is the simplest possible one,
compatible with $\kappa$-Minkowski type non-commutativity.

Let us also stress that in derivation of the  (\ref{36})--(\ref{39}) formulae
we used only the definition of covariant differentials and the
$\kappa$-Minkowski non-commutativ\-ity.
\newline

Let us now turn to the issue of definition of plane waves on
$\kappa$-Minkowski space-time, using the differential calculus
presented above. The plane waves, as in the standard case, are
defined to be fundamental solutions of an appropriate operator,
which defines standard dynamics (the non-commutative analog of the
Klein-Gordon operator).

Let us observe now that the operator $\partial$ defined in
(\ref{39}) is a natural candidate for such an operator. Indeed, it
is clearly Lorentz invariant, and is by construction closely
related to the Casimir operator (recall that $\eta_4 =
\sqrt{\kappa^2 + \eta_0^2 - \vec{\eta}\, {}^2} = \sqrt{\kappa^2 +
m^2}$.) It is now sufficient to observe that the dynamical
(deformed Klein-Gordon) operator takes the form \cite{5dcalc3}
\begin{equation}\label{43}
 0=\partial\, \Psi = \left(\partial_0^2 - \vec{\partial}\,{}^2\right)\, \Psi,
\end{equation}
where the partial derivatives are defined by eqs.~(\ref{37}), (\ref{38}). But
then it follows immediately that for the wave
\begin{equation}\label{44}
\Psi=e^{iP_0x_0}\, e^{iP_ix_i}
\end{equation}
the on-shell condition takes the form
\begin{equation}\label{45}
 \kappa^2\cosh\frac{p_0}{\kappa} - \frac{\vp}2\, e^{p_0/\kappa} =0
\end{equation}
Thus (\ref{44}) represents the on-shell massless excitation moving in the
$\kappa$-Mink\-owski space-time. Let us note at this point that one can
transform the plane wave solution (\ref{44}) to any other DSR basis, given by
${\cal P}_0, {\cal P}_i$ by simply making use of the transformation $P_0
\rightarrow P_0({\cal P}_0, {\cal P}_i)$, $P_i \rightarrow P_i({\cal P}_0,
{\cal P}_i)$.

At this point we face the following problem. In the analysis reported in
\cite{Amelino-Camelia:2002tc} the authors prove that group velocity for the
wave packet composed from the waves of the form (\ref{44})\footnote{In the
paper \cite{Amelino-Camelia:2002tc} the authors make use of the different,
Majid and Oeckl four dimensional differential calculus, which further
strengthens the result obtained here.} is $v_g = \dd{p_0}{|\vec{p}}$. Yet in
the hamiltonian analysis reported in \cite{JKM} one finds that for all DSR
theories, the velocity of massless particles equals $1$, the universal velocity
of light\footnote{This result agrees with another calculation of group velocity
of wave packets moving in $\kappa$-Minkowski space-time reported in
\cite{Tamaki}. The method  employed in this paper has  been however criticized
in \cite{Amelino-Camelia:2002tc}.}. It seems that the only way out of this
dilemma is that wave packets constructed as a linear combination of plane waves
do not represent localized particles states in the DSR theories. One should
recall at this point that  in Special Relativity the three velocity obtained
from hamiltonian analysis is given by $v_i^{(H)} = u_i/u_0$, while the group
velocity of wave packet is given by the derivative of energy with respect to
momenta calculated on-shell. However it is an accidental property of Special
Relativistic kinematics that these velocities turn out to be equal. Of course,
the proper understanding of the concept of velocity in DSR theories is the
question of central importance for the whole DSR programme and deserves urgent
studies.

\section{Relation to other formalisms}
In this section we would like to discuss relation between de Sitter space
formalism developed in the preceding sections and some other techniques used in
the context of Doubly Special Relativity. More specifically, we will discuss
the the use of classical four-momenta variables for description of
energy-momentum conservation laws \cite{lunoDSR}, \cite{Judes:2002bw} and the
method of ${\cal U}$ operator employed by Magueijo and Smolin
\cite{Magueijo:2001cr}, \cite{Magueijo:2002am}.

\subsubsection*{Classical variables variables and energy-momentum conservation}

In the papers \cite{lunoDSR},\cite{Judes:2002bw}, in order to solve the
outstanding problems of addition of momenta for multi-particle systems and
conservation laws in DSR, Lukierski and Nowicki and Judes and Visser proposed
as a simple solution to introduce auxiliary ``classical'' variables ${\cal
P}_\mu$, related to the physical momenta $p_\mu$ in a given basis in such a way
that ${\cal P}_\mu$ transform under Lorentz transformation precisely in the way
momenta in  standard Special Relativity do. The authors of these papers claim
that in order to compute total momentum of the system one must simply use the
standard rules Special Relativistic linear addition rule for ${\cal P}_\mu$,
and then just transform back the result to the physical variables.

It is a rather trivial observation that in the geometric language adopted here
the classical variables are nothing but the de Sitter coordinates $\eta_\mu$,
the latter having the required transformation rules under action of Lorentz
transformation. However, as shown in \cite{JKM}, these variables have physical
interpretation of four velocities. More specifically, if we have a point
particle carrying energy/momentum $(p_0, p_i)$ and if we compute the
four-velocity of the particle using the standard hamiltonian method, and taking
care of the non-trivial phase space structure of DSR, in {\em any} DSR theory
we find $u_\mu=\eta_\mu$. This provides a relation between four velocity (and
three velocity as well) of {\em one} particle and the energy/momentum it
carries, however a priori does n ot tell anything about total energy/momentum
of a system composed of {\em many} particles.

One should also stress that the construction of classical variables is not as
straightforward as it may apparently seem to be. The problem is that, as shown
in Section 3, the construction of these variables is not unique. There are many
functions $\eta_\mu(p_\mu)$ (or in the language of
\cite{lunoDSR},\cite{Judes:2002bw} ${\cal P}_\mu(p_\mu)$) that correspond to
given Lorentz transformation rules for $p_\mu$. It might be argued that this
freedom is in fact severely restricted, since from the results of Section 3 we
know that various choices of $\eta_\mu(p_\mu)$ are controlled by an arbitrary
function of the Casimir. However one still faces the problem, which choice
should be made for the mapping from the physical variables to the classical
ones and its inverse, in particular should both of them be the same. Moreover,
one should note that every choice of this mapping leads to different phase
space, and it seems hard to believe that energy/momentum composition law is to
be independent of the form of the phase space of the system under
consideration.

We would like to stress it once again that a DSR theory does not consist only
of the prescription of how Lorentz generators act on momenta, but of the whole
of the phase space in which dynamics of the system takes place. If DSR was a
just the momentum space Special Relativity in a nonlinear disguise, it would
not be of much interest, because, as observed in \cite{lunoDSR},
\cite{Ahluwalia:2002wf}, it most likely would be physically indistinguishable
from Special Relativity. But Doubly Special Relativity is built, as shown
above, on a highly non-trivial non-commutative space-time sector, along with
strongly deformed phase space.

To conclude, it seems that from the point of view of the geometrical analysis
presented here, the only relevance of the classical variables seem to be that
they are just identical with de Sitter variables. We will return to this issue
in the next section.

\subsubsection*{Magueijo--Smolin operator}

In order to construct their DSR theory (called nowadays DSR2) Magueijo and
Smolin \cite{Magueijo:2001cr}\footnote{The formulation of the DSR2 theory in
the de Sitter space language is presented in the Appendix.} started from the
non-linear realization of the Lorentz generators ${\cal L}$ of the form
\begin{equation}\label{31x}
{\cal L} = {\cal U}^{-1}\, L \, {\cal U},
\end{equation}
where ${\cal U}$ is some one-to-one mapping, and $L$ denotes the standard
Lorentz generators.

In the framework of the geometric de Sitter space approach to DSR,
the role of ${\cal U}$ mapping is easy to understand. This mapping
is just a one-to-one map from a subset of momentum de Sitter
space, on which the physical momenta $p_\mu$ are defined as
coordinates to a subset of the five-dimensional Minkowski space,
with coordinates $\eta_\mu$. For example, in the case of the DSR1
(\ref{26a}) we have
\begin{eqnarray}
{\eta_0} &=& {\cal U}(P_0, \vec{P}) \circ P_0 = -\kappa\, \sinh
\frac{P_0}\kappa - \frac{\vec{P}\,{}^2}{2\kappa}\,
e^{  \frac{P_0}\kappa} \nonumber\\
\eta_i &=&  {\cal U}(P_0, \vec{P}) \circ P_i= -P_i \, e^{  \frac{P_0}\kappa} .
\label{32x}
\end{eqnarray}
As in the case of the original construction presented by Magueijo and Smolin
\cite{Magueijo:2001cr} it is useful to represent ${\cal U}(p_0, \vec{p})$ as an
exponent of a linear operator acting on $p_\mu$. Then the meaning of
eq.~(\ref{31x}) is clear: one goes from de Sitter to Minkowski, performs
Lorentz transformation and that goes back to de Sitter. This is of course, a
direct counterpart of the Leibnitz rule procedure employed in Section III.

\section{Concluding remarks and open problems}

In this paper we investigated in details the de Sitter geometric picture of
Doubly Special Relativity theories. This picture is equivalent to the quantum
algebraic approach to these theories, initiated by investigations of the
$\kappa$-Poincar\'e algebra \cite{kappaP} and then generalized to incorporate
the quantum algebraic structure to all DSR theories in \cite{juse},
\cite{Kowalski-Glikman:2002jr}. In particular we show that the geometric
approach makes it possible to construct the phase space of DSR (which agrees
with the one obtained by the Heisenberg double'' prescription), the Lorentz
transformation rules for both momenta and positions, as well as the covariant
differential calculus, which is the first step in construction of the
DSR-covariant field theory.

It should be stressed that all the results reported in this paper concern
kinematics of one particle. Therefore the only physical prediction which can be
made at this stage of developments is the relation between velocity of a
particle and energy it carries. Remarkably, it turns out that for all DSR
models based on de Sitter geometry of momentum space, the speed of massless
particles equals $1$ \cite{JKM}. However, for massive particles such relation
is model dependent (and differs from the one provided by special relativity).
It is therefore very important to understand which of the DSR models describes
nature. The possible relation between DSR and the flat space, semiclassical
limit of quantum gravity, along with the experience gained by analyzing the
$2+1$ gravity models seems to suggest that the relevant basis is the classical
one. The reason is that in the natural, minimal coupling of particles to
gravity one has to do with the standard dispersion relation.

In the recent literature (see for example \cite{Ahluwalia:2002wf}) there
appeared claims that DSR is nothing but Special Relativity in non-linear
disguise. It is clear from the results presented above that this claim could
not be correct, at least in the case of our formulation of DSR, as there is no
mapping from the DSR phase space to the phase space of single particle in
special relativity. The reason is simply that the topologies of these phase
spaces are different (noncommutative $R^4$ $\times$ dS in the case of DSR
versus commutative $R^4$ $\times$ $R^4$ in SR.) Similarly there does not exist
any form of the $\kappa$-Poincar\'e Hopf algebra (which is an algebra of
symmetries of DSR) that is equivalent to the standard Poincar\'e algebra
(because the co-product of the former is non-trivial.) Thus the DSR description
of particle kinematics differs from the standard special relativistic one, and
 can be reduced to the latter only in the limit
$\kappa\rightarrow\infty$.
\newline

Many outstanding problems remain still to  be solved, of course. Let us finish
this paper with (a partial) list of the open problems that we feel are most
urgent

\begin{itemize}
\item The theory we present in this paper is to be understood as a theory of kinematics of one
particle systems. In particular we do not know yet how the proper description
of the many particle states should look like. In the course of transition from
the description one particle to many particles system one would have to define,
among others the consistent notion of the total momentum and energy, and the
conservation rules. The fact that our theory can be at best invariant under
non-commutative translations indicates that the conservation rules would differ
from the ones well known from the classical (relativistic and non-relativistic)
mechanics and field theory.
\item Keeping in mind that very sensitive time-of-flight experiments \cite{Amelino-Camelia:2002vw}
are expected to be performed in
a near future, the question of what velocity of physical particles is, is
perhaps the most urgent one in the field of DSR phenomenology.
\item The algebras we have been dealing with are understood to be commutator algebras of quantum
operators. Therefore to formulate a complete theory one investigate the
functional analysis of these operators. In particular it would be interesting
to see if these algebras lead to the appearance of the minimal length.
\item Understanding quantum mechanics on DSR phase space would make it possible to built quantum field theory with
Doubly Special Relativity playing the role analogous to the one played by
Poincar\'e symmetry in the standard QFT.
\item Last but not least, if indeed, as claimed in the Introduction, Doubly Special Relativity
can be understood as a ``Quantum Special Relativity'' the complete
understanding of DSR would without doubt be an important step in our quest for
the theory of quantum gravity.
\end{itemize}

\section*{Acknowledgement}
One of us (JKG) would like to thank Giovanni Amelino-Camelia, Achim Kempf,
Jerzy Lukierski, Joao Magueio, Lee Smolin, Jan Sobczyk  for many discussions
and comments at various stages of the work on this project. He would also like
to thank Perimeter Institute for hospitality during his stay there in December
2002, when the work on this project started.

\section*{Appendix. Another DSR bases}

In this Appendix we present the derivation of phase spaces for a class of
``classical bases'' in which momenta transform under Lorentz transformations in
the same way as in Special relativity, and some bases with transformation law
of Magueijo and Smolin.

\subsection*{The classical bases}

According to our investigations of Section 3, the most general form of $\eta$
variables in the classical basis reads
$$
 \eta_{\mu}=p_{\mu}f(p_{0}^{2}-\vp)
$$
$$
 \eta_{4}= \left(\kappa^{2}+(p_{0}^{2}-\vp)f^2(p_{0}^{2}-\vp)\right)^{1/2} \equiv \sqrt{\kappa^{2}+ m^2\, f^2},
$$
where we used abbreviation $m^2 \equiv p_{0}^{2}-\vp$. The
functions $f$ are restricted only by the requirement that in the
limit $\kappa\rightarrow\infty$, $f\rightarrow1$.

Let us assume now that
$$
 [x_{0},p_{i}]=ip_{i}A,\quad [x_{0},p_{0}]= iB
$$
where $A$ and $B$ are rotational invariant functions. It follows from
$$
 [x_{0},\eta_{i}]=0
$$
that
\begin{equation}\label{a.1}
  2f'p_0\, B +  A(f -2\vp\, f') =0,
\end{equation}
where $f'$ denotes derivative of $f$ with respect to its argument.
 From
$$
 [x_{0},\eta_{0}]=\frac{i}{\kappa}\eta_{4}
$$
we find
\begin{equation}\label{a.2}
(f+2p^2_{0}f')\,B
 -2p_{0}\vp\,f'\,A
 =  \frac{1}{\kappa}\sqrt{\kappa^{2}+ m^2\, f^2}
\end{equation}
Note that equations (\ref{a.1}), (\ref{a.2}) become degenerate for
$f'=0$ (the equation $f -2\vp\, f' =0$ does not have any
solutions.)  With no loss of generality we can assume that  $f=
1$. Below we will consider the degenerate case, and the generic
non-degenerate one.
\newline

In the case $f=1$ we have to do simply with the algebra (\ref{10}), (\ref{11})
with $\eta_4$ replaced with $\sqrt{\kappa^2 +p_0^2-\vp}$:
\begin{equation}\label{a.3}
   [x_0,p_0] = \frac{i}\kappa\, \sqrt{\kappa^2 +p_0^2-\vp}, \quad [x_0,p_i] = 0,
\end{equation}
\begin{equation}\label{a.4}
   [x_i, p_0] =\frac{i}\kappa\, p_i, \quad [x_i, p_j] = \frac{i}\kappa\,
\delta_{ij}(p_0 - \sqrt{\kappa^2 +p_0^2-\vp}),
\end{equation}

In the generic case, we have
\begin{equation}\label{a.5}
 [x_{0},p_{0}]=\frac{i}{\kappa}\frac{\left(\kappa^{2}+m^2\, f^{2}\right)^{1/2}
 (f+2\vp\frac{\partial f}{\partial \vp})}{f\left(f-2m^2\,
 \frac{\partial f}{\partial \vp}\right)}
\end{equation}
\begin{equation}\label{a.6}
 [x_0, p_i] = i p_i\, \frac{1}{\kappa}\frac{\left(\kappa^{2}+m^2\, f^{2}\right)^{1/2}
 2p_{0}\frac{\partial f}{\partial \vp}}{f\left(f-2m^2\,
 \frac{\partial f}{\partial \vp}\right)}
\end{equation}
The remaining commutators have the form
\begin{equation}\label{a.7}
 [x_{i},p_{j}]=i(\delta_{ij}B+p_{i}p_{j}C), \quad [x_{i},p_{0}]=ip_{i}(\frac{1}{\kappa}+p_{0}C)
\end{equation}
where
$$
 B=\frac{1}{\kappa}\frac{p_{0}f-
 \left(\kappa^{2}+m^2\, f^{2}\right)^{1/2}}{f},
$$
$$
 C=\frac{1}{\kappa}\frac{2\frac{\partial f}{\partial \vp}(p_{0}-\kappa B)}
 {f-2m^2\, \frac{\partial f}{\partial \vp}},
$$
where, as above $m^2 \equiv p_{0}^{2}-\vp$.

\subsection*{Magueio--Smolin basis (DSR2)}

The Magueio--Smolin basis (DSR2) \cite{Magueijo:2001cr} is defined
by
\begin{equation}
 \eta_{\mu}=\frac{P_{\mu}}{1-P_{0}/\kappa}
\end{equation}
Let us derive the phase space commutators. By making use of
$$
 [x_{0},\eta_{0}]=\frac{i}{\kappa}\eta_{4}
$$
we find
\begin{equation}
 [x_{0},P_{0}]=\frac{i}{\kappa}(\kappa^{2}(1-P_{0}/\kappa)^{2}+P_{0}^{2}-\vec{P}^{2})^{1/2}(1-P_{0}/\kappa)
\end{equation}
and from
$$
 [x_{0},\eta_{i}]=0
$$
we obtain
\begin{equation}
 [x_{0},P_{i}]=-\frac{i}{\kappa}(\kappa^{2}(1-P_{0}/\kappa)^{2}+P_{0}^{2}-\vec{P}^{2})^{1/2}.
\end{equation}
Next, from
$$
 [x_{i},\eta_{0}]=\frac{i}{\kappa}\eta_{i}
$$
it follows that
\begin{equation}
 [x_{i},P_{0}]=iP_{i}(1-P_{0}/\kappa),
\end{equation}
while from
$$
 [x_{i},\eta_{j}]=\frac{i}{\kappa}\delta_{ij}(\eta_{0}-\eta_{4})
$$
\begin{equation}
 [x_{i},P_{j}]=\frac{i}{\kappa}\left(\delta_{ij}\left(P_{0}-(\kappa^{2}(1-P_{0}/\kappa)^{2}+P_{0}^{2}-\vec{P}^{2})^{1/2}\right)
 -P_{i}P_{j} \right)
\end{equation}
 In the paper \cite{Kowalski-Glikman:2002jr} we made use of the
 another form of the DSR2 basis, given by
 \begin{equation}
 \eta_{0}=-p_{0}\left(1-\frac{2p_{0}}{\kappa}+\frac{\vec{p}^{2}}
 {\kappa^{2}} \right)^{-1/2}
\end{equation}
\begin{equation}
 \eta_{4}=(\kappa-p_{0})\left(1-\frac{2p_{0}}{\kappa}+\frac{\vec{p}^{2}}
 {\kappa^{2}} \right)^{-1/2}
\end{equation}
\begin{equation}
 \eta_{i}=-p_{i}\left(1-\frac{2p_{0}}{\kappa}+\frac{\vec{p}^{2}}
 {\kappa^{2}} \right)^{-1/2}
\end{equation}
with the phase space
\begin{equation}
 [p_0, x_i] =  -\frac{i}\kappa p_i
\end{equation}
\begin{equation}
 [p_0, x_0] = i\left(1 - \frac{2p_0}\kappa \right)
\end{equation}
\begin{equation}
  [p_i, x_j] = -i \, \delta_{ij}
\end{equation}
\begin{equation}
 [p_i, x_0] = -\frac{i}\kappa\, P_i.
\end{equation}
The relation between these bases reads
\begin{equation}
 P_{\mu}=-\frac{p_{\mu}}{(1-2p_{0}/\kappa
 +\vec{p}^{\;2}/\kappa^2)^{1/2}-p_{0}/\kappa}.
\end{equation}


\begin{thebibliography}{99}
%\cite{Amelino-Camelia:2002vw}
\bibitem{Amelino-Camelia:2002vw}
G.~Amelino-Camelia, ``Quantum-gravity phenomenology: Status and prospects,''
Mod.\ Phys.\ Lett.\ A {\bf 17}, 899 (2002) [arXiv:gr-qc/0204051].
%%CITATION = GR-QC 0204051;%%

%\cite{Amelino-Camelia:2000ge}
\bibitem{Amelino-Camelia:2000ge}
G.~Amelino-Camelia, ``Testable scenario for relativity with minimum-length,''
Phys.\ Lett.\ B {\bf 510}, 255 (2001) [arXiv:hep-th/0012238].
%%CITATION = HEP-TH 0012238;%%

%\cite{Amelino-Camelia:2000mn}
\bibitem{Amelino-Camelia:2000mn}
G.~Amelino-Camelia, ``Relativity in space-times with short-distance structure
governed by an observer-independent (Planckian) length scale,'' Int.\ J.\ Mod.\
Phys.\ D {\bf 11}, 35 (2002) [arXiv:gr-qc/0012051].
%%CITATION = GR-QC 0012051;%%

\bibitem{add} Luis J.\ Garay, Int.J.Mod.Phys. {\bf A10} (1995) 145-166; T.\ Padmanabhan,
Class.Quant.Grav. {\bf 4} (1987) L107.


%\cite{Ahluwalia:2002ye}
\bibitem{Ahluwalia:2002ye}
D.~V.~Ahluwalia, ``Fermions, bosons, and locality in special relativity with
two invariant  scales,'' arXiv:gr-qc/0207004.
%%CITATION = GR-QC 0207004;%%

%\cite{Schuller:2002fn}
\bibitem{Schuller:2002fn}
F.~P.~Schuller, ``Born-Infeld kinematics,'' Annals Phys.\  {\bf 299} (2002) 174
[arXiv:hep-th/0203079].
%%CITATION = HEP-TH 0203079;%%

%\cite{KalyanaRama:2002qh}
\bibitem{KalyanaRama:2002qh}
S.~Kalyana Rama, ``Classical velocity in kappa-deformed Poincare algebra and a
maximum acceleration,'' Mod.\ Phys.\ Lett.\ A {\bf 18} (2003) 527
[arXiv:hep-th/0209129].
%%CITATION = HEP-TH 0209129;%%


%\cite{Feoli:2002dv}
\bibitem{Feoli:2002dv}
A.~Feoli, ``Maximal acceleration or maximal accelerations?,'' Int.\ J.\ Mod.\
Phys.\ D {\bf 12} (2003) 271 [arXiv:gr-qc/0210038].
%%CITATION = GR-QC 0210038;%%

%\cite{Freidel:2003sp}
\bibitem{Freidel:2003sp}
L.~Freidel, J.~Kowalski-Glikman and L.~Smolin, ``2+1 gravity and doubly special
relativity,'' arXiv:hep-th/0307085.
%%CITATION = HEP-TH 0307085;%%


%\cite{Amelino-Camelia:2003xp}
\bibitem{Amelino-Camelia:2003xp}
G.~Amelino-Camelia, L.~Smolin and A.~Starodubtsev, ``Quantum symmetry, the
cosmological constant and Planck scale  phenomenology,'' arXiv:hep-th/0306134.
%%CITATION = HEP-TH 0306134;%%


\bibitem{jkgminl} J.~Kowalski-Glikman,
``Observer independent quantum of mass,'' Phys.\ Lett.\ A {\bf 286} (2001) 391
[arXiv:hep-th/0102098].

\bibitem{rbgacjkg} N.~R.~Bruno, G.~Amelino-Camelia and J.~Kowalski-Glikman,
``Deformed boost transformations that saturate at the Planck scale,'' Phys.\
Lett.\ B {\bf 522} (2001) 133 [arXiv:hep-th/0107039].

\bibitem{kappaP} J.~Lukierski, H.~Ruegg, A.~Nowicki and V.~N.~Tolstoi,
``Q deformation of Poincare algebra,'' Phys.\ Lett.\ B {\bf 264} (1991) 331.
\bibitem{kappaM} S.~Majid and H.~Ruegg, ``Bicrossproduct structure of kappa Poincare group
and noncommutative geometry,'' Phys.\ Lett.\ B {\bf 334} (1994) 348
[arXiv:hep-th/9405107]; J.~Lukierski, H.~Ruegg and W.~J.~Zakrzewski,
``Classical quantum mechanics of free kappa relativistic systems,'' Annals
Phys.\  {\bf 243} (1995) 90 [arXiv:hep-th/9312153].

%\cite{Magueijo:2001cr}
\bibitem{Magueijo:2001cr}
J.~Magueijo and L.~Smolin, ``Lorentz invariance with an invariant energy
scale,'' Phys.\ Rev.\ Lett.\  {\bf 88} (2002) 190403 [arXiv:hep-th/0112090].
%%CITATION = HEP-TH 0112090;%

\bibitem{juse} J.~Kowalski-Glikman and S.~Nowak,
``Doubly special relativity theories as different bases of kappa-Poincare
algebra,'' Phys.\ Lett.\ B {\bf 539} (2002) 126 [arXiv:hep-th/0203040].

%\cite{Kowalski-Glikman:2002jr}
\bibitem{Kowalski-Glikman:2002jr}
J.~Kowalski-Glikman and S.~Nowak, ``Non-commutative space-time of doubly
special relativity theories,'' Int.\ J.\ Mod.\ Phys.\ D {\bf 12} (2003) 299
[arXiv:hep-th/0204245].
%%CITATION = HEP-TH 0204245;%%

\bibitem{crossalg}  A.~Nowicki, {\tt math.QA/9803064}.

\bibitem{luno} J. Lukierski and A. Nowicki, Proceedings of
Quantum Group Symposium at Group 21, (July 1996, Goslar) Eds. H.-D. Doebner and
V.K. Dobrev, Heron Press, Sofia, 1997, p. 186.

\bibitem{snyder} H.~S.~Snyder,
``Quantized Space-Time,'' Phys.\ Rev.\  {\bf 71} (1947) 38.

%\cite{Kowalski-Glikman:2002ft}
\bibitem{Kowalski-Glikman:2002ft}
J.~Kowalski-Glikman, ``De Sitter space as an arena for doubly special
relativity,'' Phys.\ Lett.\ B {\bf 547} (2002) 291 [arXiv:hep-th/0207279].
%%CITATION = HEP-TH 0207279;%%

\bibitem{lunoDSR} J.~Lukierski and A.~Nowicki,
``Doubly Special Relativity versus $\kappa$-deformation of
relativistic kinematics,'' Int.\ J.\ Mod.\ Phys.\ A {\bf 18}
(2003) 7 [arXiv:hep-th/0203065].




\bibitem{Magueijo:2002am}
J.~Magueijo and L.~Smolin, ``Generalized Lorentz invariance with an invariant
energy scale,'' arXiv:gr-qc/0207085.




\bibitem{5dcalc1} A.~Sitarz, ``Noncommutative differential calculus on the kappa
Minkowski space,'' Phys.\ Lett.\ B {\bf 349} (1995) 42 [arXiv:hep-th/9409014].
\bibitem{5dcalc2} J.A.~de~Azcarraga and J.C.~P\'erez Bueno, ``Relativistic and Newtonian
$\kappa$-spacetimes'', J.~Math.~Phys. {\bf 36} (1995) 6879
[arXive:q-alg/9505004]; P.~Kosinski, P.~Maslanka, and J.~Sobczyk, ``The
bicovariant differential calculus on the $\kappa$-Poincar\'e group and on the
$\kappa$-Minkowski space'' [arXive:q-alg/9508021].
\bibitem{5dcalc3} S.~Giller, C.~Gonera,
P.~Kosinski, and P.~Maslanka, ``A note on geometry of $\kappa$-Minkowski
space'', Acta Phys.\ Polon.\ B {\bf 27} (1996) 2171 [arXive:q-alg/9602006];
P.~Kosinski, and P.~Maslanka, J.~Lukierski, and A.~Sitarz, ``Towards
$\kappa$-deformed D=4 relativistic field theory'', Chech.\ J.\ Phys.\ {\bf 48}
(1998) 1407.

\bibitem{majid} R.~Oeckl, ``Classification of differential calculi on $U_q(b_+)$, classical limit and duality''
 J. Math. Phys. {\bf 40} (1999) 3588-3604  [arXiv:math.QA/9807097]; S.~Majid, {\em
Foundation of Quantum Group Theory}, Cambridge University Press, 1995.
\bibitem{JKM} M.\ Daszkiewicz, K.\
Imilkowska, and J.\ Kowalski--Glikman, ``Velocity of particles in
Doubly Special Relativity'', arXiv:hep-th/0304027.
%\cite{Amelino-Camelia:2002tc}
\bibitem{Amelino-Camelia:2002tc}
G.~Amelino-Camelia, F.~D'Andrea and G.~Mandanici, ``Group velocity in
noncommutative spacetime,'' arXiv:hep-th/0211022.
%%CITATION = HEP-TH 0211022;%%
\bibitem{Tamaki} T.\ Tamaki, T.\ Harada, U. Miyamoto, and T.\ Torii,
``Have we already detected astrophysical symptoms of space-time
non-commutativity?'', Phys.\ Rev.\ D {\bf 65} (2002) 083003
[arXive:gr-qc/0111056].

%\cite{Judes:2002bw}
\bibitem{Judes:2002bw}
S.~Judes and M.~Visser ``Conservation Laws in Doubly Special Relativity,''
arXiv:gr-qc/0205067.
%%CITATION = GR-QC 0205067;%%

%\cite{Ahluwalia:2002wf}
\bibitem{Ahluwalia:2002wf}
D.~V.~Ahluwalia--Khalilova, ``Operational indistinguishabilty of doubly special
relativities from  special relativity,'' arXiv:gr-qc/0212128.


\end{thebibliography}
\end{document}